\documentclass[final,3p,times,twocolumn]{elsarticle}
\biboptions{sort&compress}
\usepackage{graphicx,color}
\usepackage{latexsym,amssymb}
\usepackage{amsmath}

\def\be{\begin{equation}}
\def\ee{\end{equation}}
\def\bc{\begin{center}}
\def\ec{\end{center}}

\def\calh{{\cal H}}

\def\cali{{\cal J}}
\def\phi{\varphi}

\journal{Optics Communications}

\begin{document}
\begin{frontmatter}

\title{Photonic crystal with left-handed components}
\author[rvt]{Peter  Marko\v{s}}
\ead{peter.markos@fmph.uniba.sk}
\address[rvt]{%
Dept. Experimental Physics, Faculty of Mathematics, Physics and Informatics, 
Comenius University in Bratislava.
Slovakia}

\begin{abstract}
We show that the periodic array of left-handed cylinders  possesses a rich spectrum of guided modes
when the negative permeability of cylinders equals exactly to minus value of permeability  of embedding media. 
These resonances strongly influences propagation of electromagnetic waves through photonic structures made from left-handed materials. 
A series of Fano resonances excited by incident wave destroys the band frequency  spectrum of
square array of left-handed  cylinders and increases considerably the absorption  of transmitted waves.

\end{abstract}


\end{frontmatter}

\section{Introduction}

Left-handed (LH) materials which   possess, in certain frequency interval,
simultaneously negative electric permittivity $\varepsilon<0$ 
and magnetic permeability $\mu<0$  exhibit interesting  physical and optical properties, not observed in standard
dielectrics or metals \cite{veselago,phys-world,ez}. 
Although LH materials are not commonly available in nature, they may be prepared in laboratories and used for construction 
of devices with prescribed optical properties. 

Physical and optical properties of LH materials have been  studied during last 15 years from two  different points of view. The  first, microscopic, approach  
concentrates on the analysis of   the design of 
individual ``atoms'' from which periodic macroscopic LH structure is constructed. 
The aim of this research  is 
to optimize the structure of individual ``atom''  with respect to its resonant response which  guarantee 
required  properties of resulting  macroscopic
material \cite{11,pendry-2000,science,abs}. The second, macroscopic, approach considers  homogeneous  LH medium 
with negative $\varepsilon$ and $\mu$ and investigates its physical properties as well as  
possible application 
 of LH material in various  photonic devices.  Typical  example of these  studies  are 
detailed theoretical and numerical  investigations  of  one dimensional layered structures composed from alternating LH and dielectric layers 
\cite{li,rpw,1d-gred,mh}.

Recently it has been shown that  
components made from LH materials may strongly influence 
optical properties of 2D photonic structures and lead to unexpected  phenomena, not observable in conventional photonic structures
made from dielectric materials.
For instance,  the square periodic array of cylinders made from LH material can possess a non-standard band frequency  
spectrum which contains
the so-called  folded bands \cite{chen,busch}. 
Such results motivates further investigation  how the application of LH materials  influences the functionality of optical composites.

\begin{figure}[t]
\begin{center}
\includegraphics[width=0.16\textwidth]{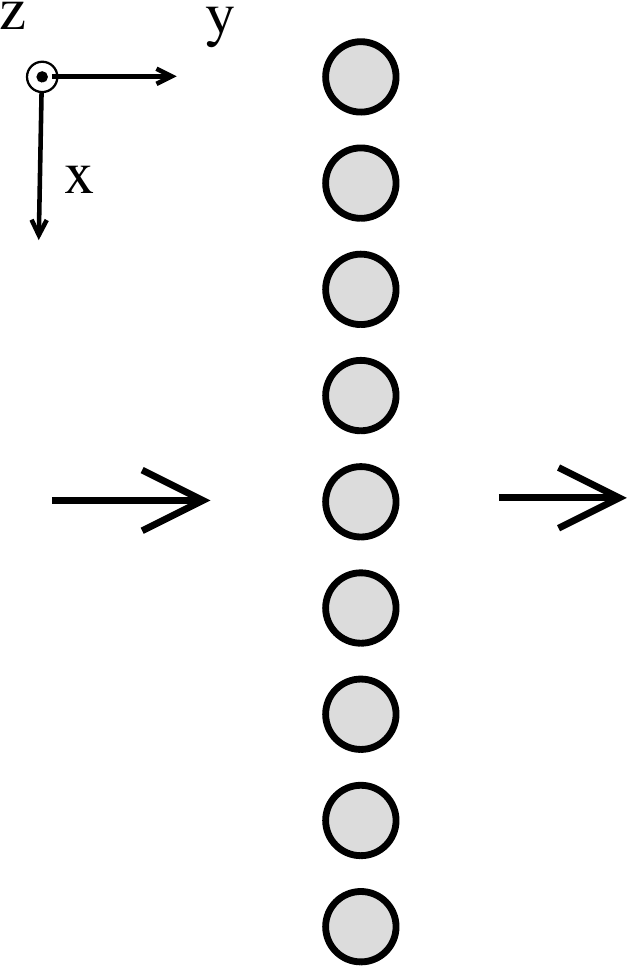}
~~~~
\includegraphics[width=0.25\textwidth]{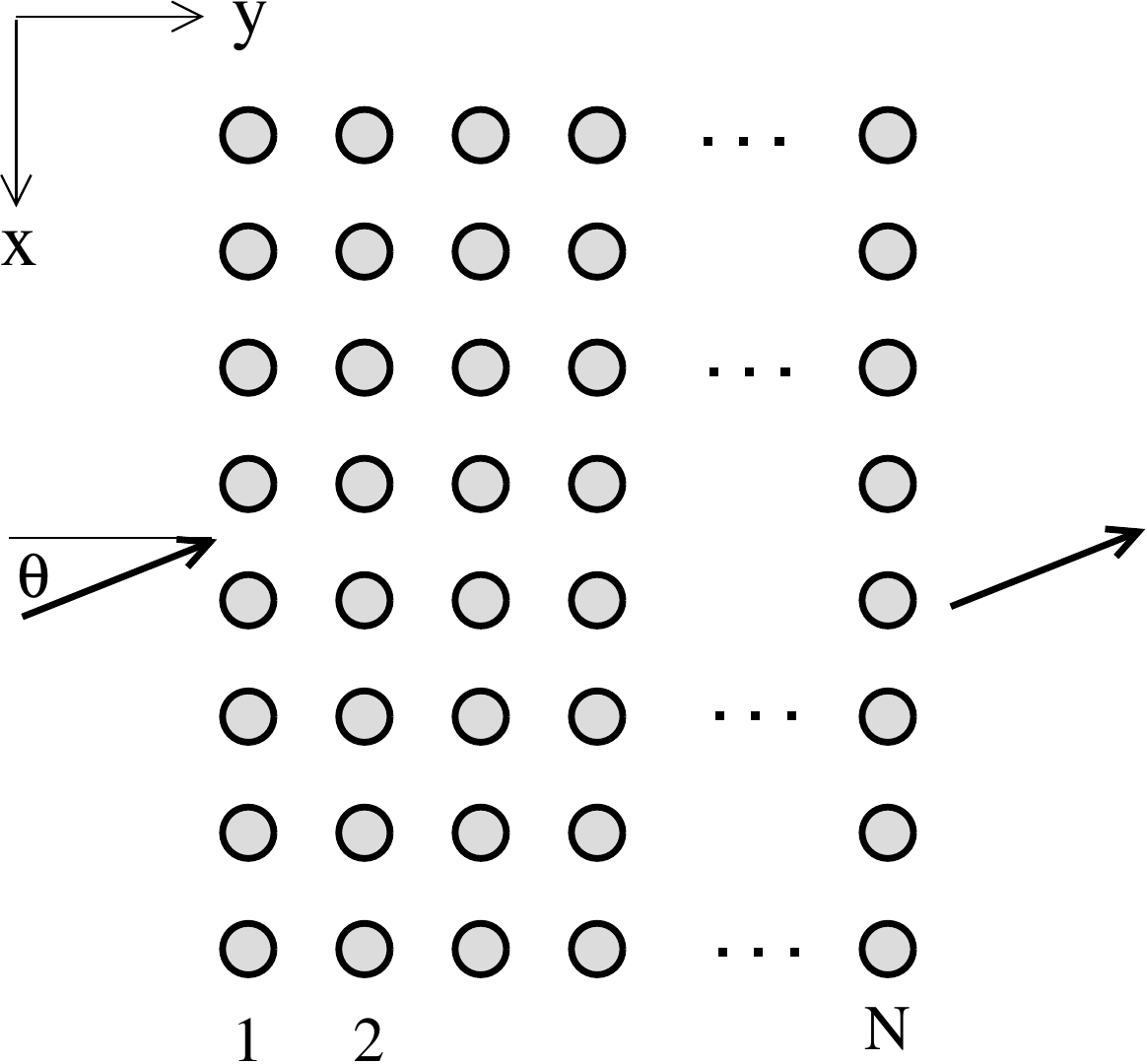}

\footnotesize{
(a)~~~~~~~~~~~~~~~~~~~~~~~~~~~~~~~~~~~~~~~~(b)~~~~~~~~~~~~~~~~~~~~~~~~~~~~~~
}
\caption{(a) Periodic linear chain of homogeneous cylinders.
Cylinders are parallel to the $z$ axis, their radius is $R$,  permittivity $\varepsilon = -12 $ and permeability $\mu = -1$.
The entire structure is periodic in the $x$ direction with spatial period $a$ which is used as a length unit throughout this paper.
(b) $N$ parallel chains of cylinders located in planes $y= na$, $0\le n\le N-1$.
The embedding medium is vacuum with permittivity $\varepsilon_1=1$ and permeability $\mu_1=1$.
The incident electromagnetic  wave propagating  along  the $y$ direction with  either   $E_z$ ($E\parallel z$)  or $H_z$ ($H\parallel z$) polarization  
has the wavelength $\lambda$ and 
dimensionless frequency
$f=a/\lambda$.
}
\label{fig:geom}
\end{center}
\end{figure}

\begin{figure}[t]

\begin{center}
\includegraphics[width=0.32\textwidth]{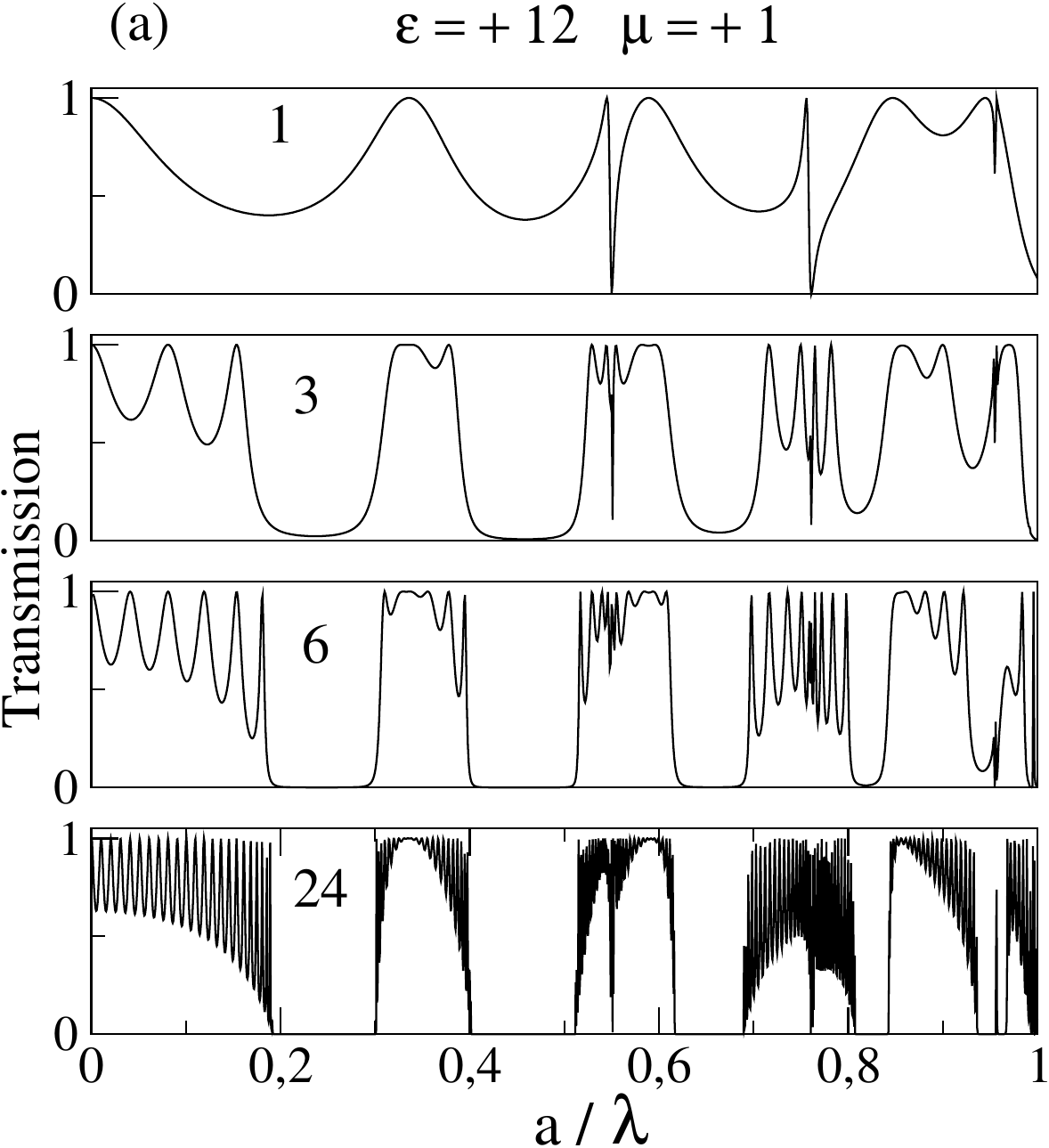}\\
\includegraphics[width=0.32\textwidth]{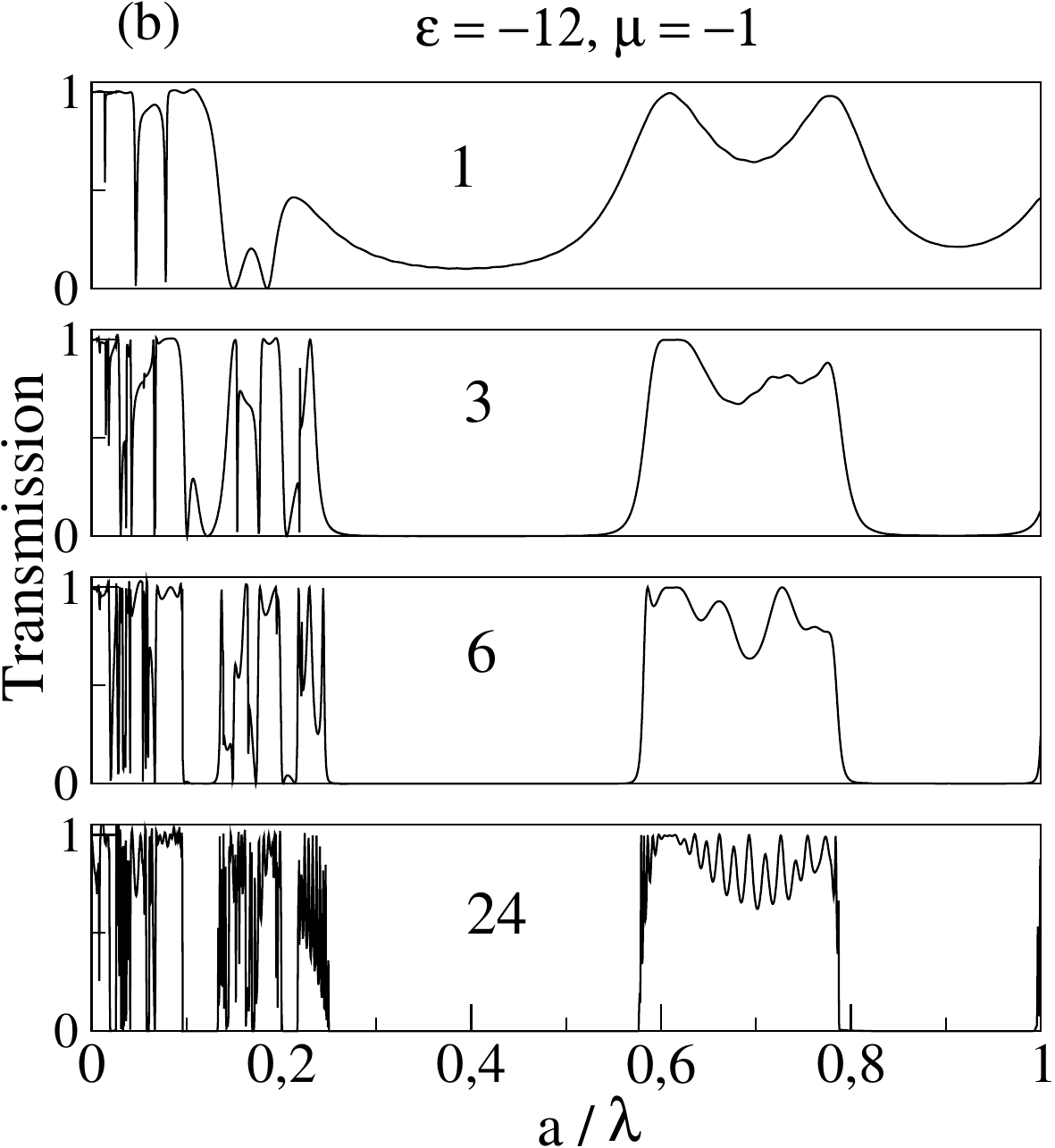}
\caption{
Transmission spectra for $E_z$-polarized electromagnetic wave propagating through  arrays of (a) dielectric and (b) LH cylinders. 
The spectra were calculated for $N=1$, 3, 6 and 24 rows of cylinders shown in Fig. \ref{fig:geom}.
The cylinder radius $R=0.3a$.
For dielectric cylinders, the spectrum evolves to transmission  bands and gaps when the number of rows increases
\cite{WP}. 
In contrast, for  LH
cylinders,  the typical  band structure arises  only for sufficiently  high frequencies,  $a/\lambda > 0.2$. For smaller  frequencies, the 
transmission spectrum consists from highly irregular  series of maxims and minims.
}
\label{fig:ukazka}
\end{center}
\end{figure}

In this Paper we discuss  another unusual  property of LH photonic structures.
We consider  periodic arrangements of LH  cylinders shown in Fig. \ref{fig:geom} and calculate  numerically
the frequency dependence of the transmission coefficient of incident electromagnetic (EM) wave.
Instead of regular transmission bands and gaps, typical for spectra  of spatially  periodic dielectric 
structures \cite{joan-pc,sakoda},  we observe, for small frequencies,  a series of irregular maxims and minims in the frequency dependence of transmission coefficient. 
An example of such irregular frequency dependence is shown in Fig. \ref{fig:ukazka}(b). 
We show that physical origin of these irregularities lies in the   excitation of high number of leaky eigenmodes of the structure and consequent interference of 
excited field with incident electromagnetic field
\cite{fan,sfan,rybin}.  Similar resonances 
were observed  previously in   dielectric photonic structures  
\cite{on}
and their influence on  the optical response has been  analyzed in Refs.
\cite{fano,luk,pm-pra}. 
We found that  the spectrum of excited resonances strongly depends on actual values of
negative permittivity and permeability
and is extremely rich if one of these parameters coincides with the minus value of the corresponding parameter of embedding media.

The Paper is organized as follows. In Sect. 2, we calculate
transmission coefficient of perpendicularly incident EM wave propagating  across  the 
linear array  of cylinders and  through slab constructed from finite number of parallel  arrays. 
We identify a series of resonances  in the frequency dependence of transmission coefficient.
To find their physical interpretation,
we analyze the complete spectrum of guided modes  of the linear chain of cylinders and 
show that 
maxima  and minima of the transmission coefficient are associated with Fano resonances excited in the photonic structure by incident EM wave.
In Section 3 we study how the electromagnetic response of the structure  depends on  material parameters: radius of cylinders,   
magnetic permeability and absorption. Of particular interest is the model with frequency dependent negative permittivity and permeability  which also exhibit 
a series of resonances in the transmission spectra in the  frequency interval where either permittivity or permeability approaches the value -1. Since the the resonance causes a strong enhancement of electromagnetic field  inside the structure, we calculate the absorption of transmitted wave.
Conclusion is given in Sect. 4. Finally,
numerical method used for the calculation of the transmissions coefficient is described   in Appendices.


\section{Linear chain of left-handed cylinders}

In this Section, we study the response of
linear chain of LH cylinders to incident electromagnetic wave.  
The structure, shown in Fig. \ref{fig:geom}(a), consists from an infinite periodic  chain of cylinders
embedded in the vacuum with permittivity $\varepsilon=+1$ and permeability $\mu=+1$.
The spatial periodicity of the structure along the $x$ direction,  given by
distance $a$ between neighboring cylinders, defines the length unit. 
Cylinders  are  infinite along the $z$ direction and 
are made from homogeneous material with 
relative permittivity $\varepsilon=-12$  and permeability $\mu=-1$. 

Incident plane wave with wavelength $\lambda$ propagates along the $y$ direction. We calculate the transmission  and reflection coefficient
as a function of dimensionless frequency $f = a/\lambda$. The method is described in Appendix A. Here we only note that
electromagnetic field is expanded in series of cylinder functions \cite{stratton} given by Eqs. 
(\ref{eq:inc}) and  (\ref{eq:onc}).
Expansion coefficients $\alpha$ and $\beta$ are calculated from the requirement of continuity of tangential components
of electric and magnetic intensity at the boundary of cylinders.

\begin{figure}[t]
\centering
{\includegraphics[width=0.98\linewidth]{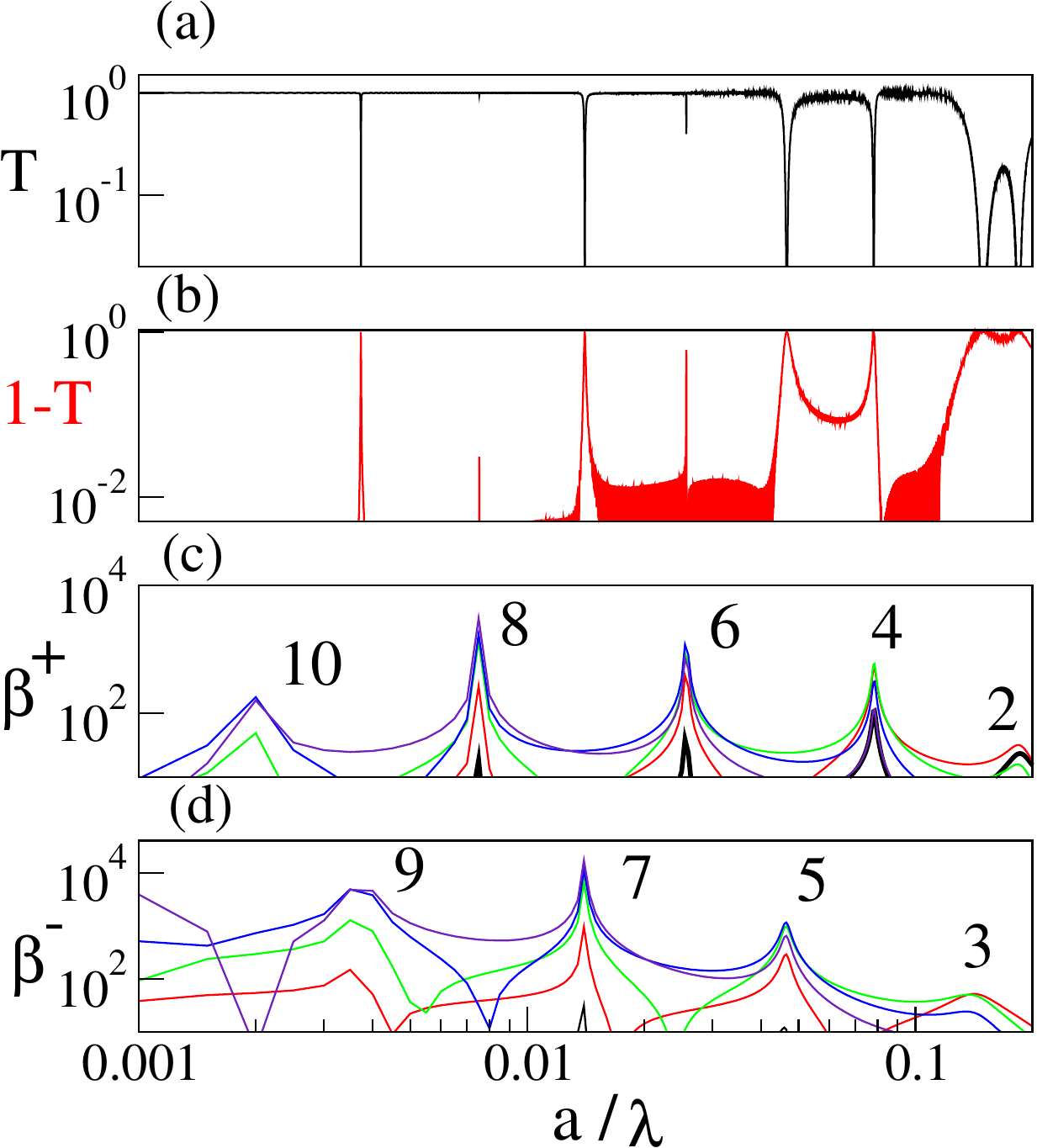}}
\caption{Color online) 
(a) The frequency dependence of the transmission coefficient of EM wave propagating through linear chain of LH cylinders.
Note logarithmic scale on horizontal axis which is identical for all panels.
We plot in panel (b) 
also the difference  $1-T$
which enables us to identify very tiny Fano resonances, not visible in panel (a).
Two bottom panels  present resonant frequency dependence of 
 coefficients $\beta^+_{2l}$ (c) and  $\beta^-_{2l+1}$ (d) defined by Eq. (\ref{eq:onc}). 
For numerical reasons, we plot a ratio  $\beta^\pm_k/H'_k(2\pi R/\lambda$), 
where $H'_k$ is a derivative of the first Hankel function.
Since incident field is symmetric with respect to the transformation $x\to -x$, 
only spatially symmetric  resonances are excited \cite{pc-asym,pm-pra}.
The intensity of electric field excited inside cylinders is displayed in Fig. \ref{eps-12-pole}.
}
\label{eps-12-bb}
\end{figure}

The frequency dependence of the transmission coefficient of the $E_z$-polarized 
 plane EM wave for both dielectric and LH cylinders 
is  shown in top panels of Figs. \ref{fig:ukazka}(a,b). 
For dielectric cylinders, we identify three deep minima in $T$ which correspond to Fano resonances excited by incident wave \cite{fan,pm-pra}.
Similar resonances were found  in the LH structure, but their number is much higher and resonant frequencies  are located in the region of small
frequencies $a/\lambda < 0.2$. Detailed frequency dependence of the transmission coefficient is given in Fig.~\ref{eps-12-bb}.

Lower panels of Figs. \ref{fig:ukazka}(a,b) display the transmission coefficient for
photonic slabs composed from 3, 6 and 24 rows of cylinders.  
For dielectric cylinders, Fano resonances develop to narrow Fano bands discussed in details in Ref. \cite{pm-pra}.
In contrast, a high number of Fano resonances  in the  LH structure,
completely destroy the band structure.
Regular transmission band, typical for periodic media, is observed only for higher frequencies
 $a/\lambda\approx 0.58$.

\begin{figure}[t]
\begin{center}
\includegraphics[width=0.29\linewidth]{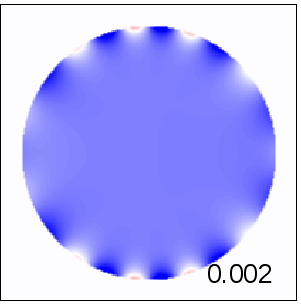}
~~~
\includegraphics[width=0.29\linewidth]{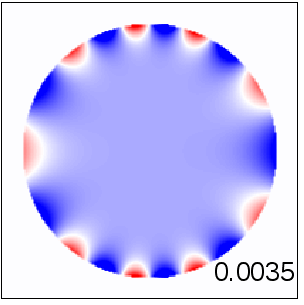}
~~~
\includegraphics[width=0.29\linewidth]{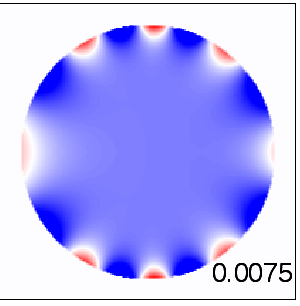}\\
\includegraphics[width=0.29\linewidth]{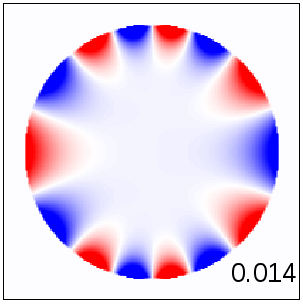}
~~~
\includegraphics[width=0.29\linewidth]{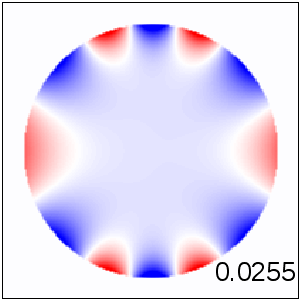}
~~~
\includegraphics[width=0.29\linewidth]{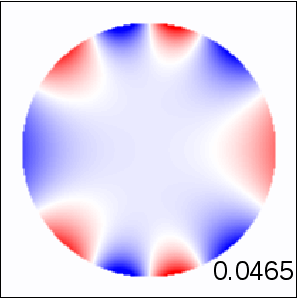}\\
\includegraphics[width=0.29\linewidth]{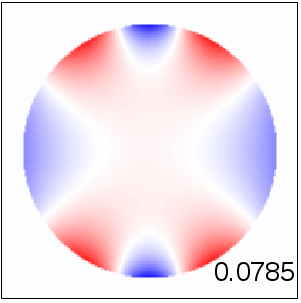}
~~~
\includegraphics[width=0.29\linewidth]{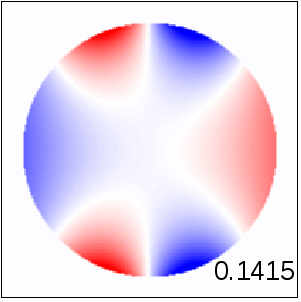}
~~~
\includegraphics[width=0.29\linewidth]{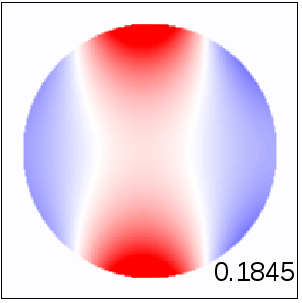}
\footnotesize{
~~\\
$\longrightarrow$  propagation of $E_z$ polarized EM wave
}
\end{center}
\caption{(Color online) 
The spatial distribution of  electric intensity $e_z$ inside cylinders 
arranged in a linear chain for resonances $2\le k\le 10$.
 Resonant dimensionless frequencies $a/\lambda$, identified in Fig. \ref{eps-12-bb}, are given in legends.  Index $k$ determines the period of the wave around the cylinder surface in agreement with Eq. (\ref{kk}).
The incident $e_z$ polarized electromagnetic wave propagates from left to the right (Fig.~\ref{fig:geom}).
}
\label{eps-12-pole}
\end{figure}


Following Ref. \cite{pm-pra} we expect that the strong irregularities in  transmission spectra of linear chain of cylinders 
 are caused by excitation of resonances  in the structure. 
To verify this conjecture, we calculate  how coefficients
$\beta$, which measure the amplitude of excited resonances (Eq. \ref{eq:onc}) 
depend on the frequency of incident wave.
We found, in agreement with Fig. \ref{eps-12-bb}(a),
a series of strong resonances, associated with sharp narrow maxims in $\beta$ (Fig. \ref{eps-12-bb}(c,d)).
Resonant frequencies coincides  with positions of minims in transmission coefficient.
Note that owing to  even symmetry of perpendicularly incident plane EM wave,
only resonances symmetric with respect to transformation $x\to -x$ were excited \cite{pc-asym}.
Resonant frequencies  lye very close to each other (note logarithmic scale of horizontal axis). 
Comparison of panels (a,b) with (b,c) confirms that each observed transmission minimum corresponds to  Fano resonant frequencies. 

Data shown in Fig. \ref{eps-12-bb} enable us to identify  exponential decrease of  resonant frequency $f_k$ when
  $k$ increases:
\be\label{kk}
f_{k} = f_0 e^{ - ck}
\ee
with constant $c\approx 0.58$.

Figure \ref{eps-12-pole} displays spatial 
distribution of  electric field $e_z$ inside LH cylinders for nine resonant  frequencies $2\le k\le 10$ identified in Fig. \ref{eps-12-bb}.
The field  concentrated close to the  boundary of cylinders. 
The spatial distribution of field along the boundary reminds us the circular standing waves with the wavelength
\begin{equation}
\label{eq:lambda}
\Lambda = \frac{2\pi }{k}~R.
\end{equation}
Similar inhomogeneous spatial distribution of EM field has been found also in other photonic systems which contains  LH cylinders \cite{aaa}.

\subsection{Eigenmodes}

\begin{figure}[t]
\centering
{\includegraphics[width=0.9\linewidth]{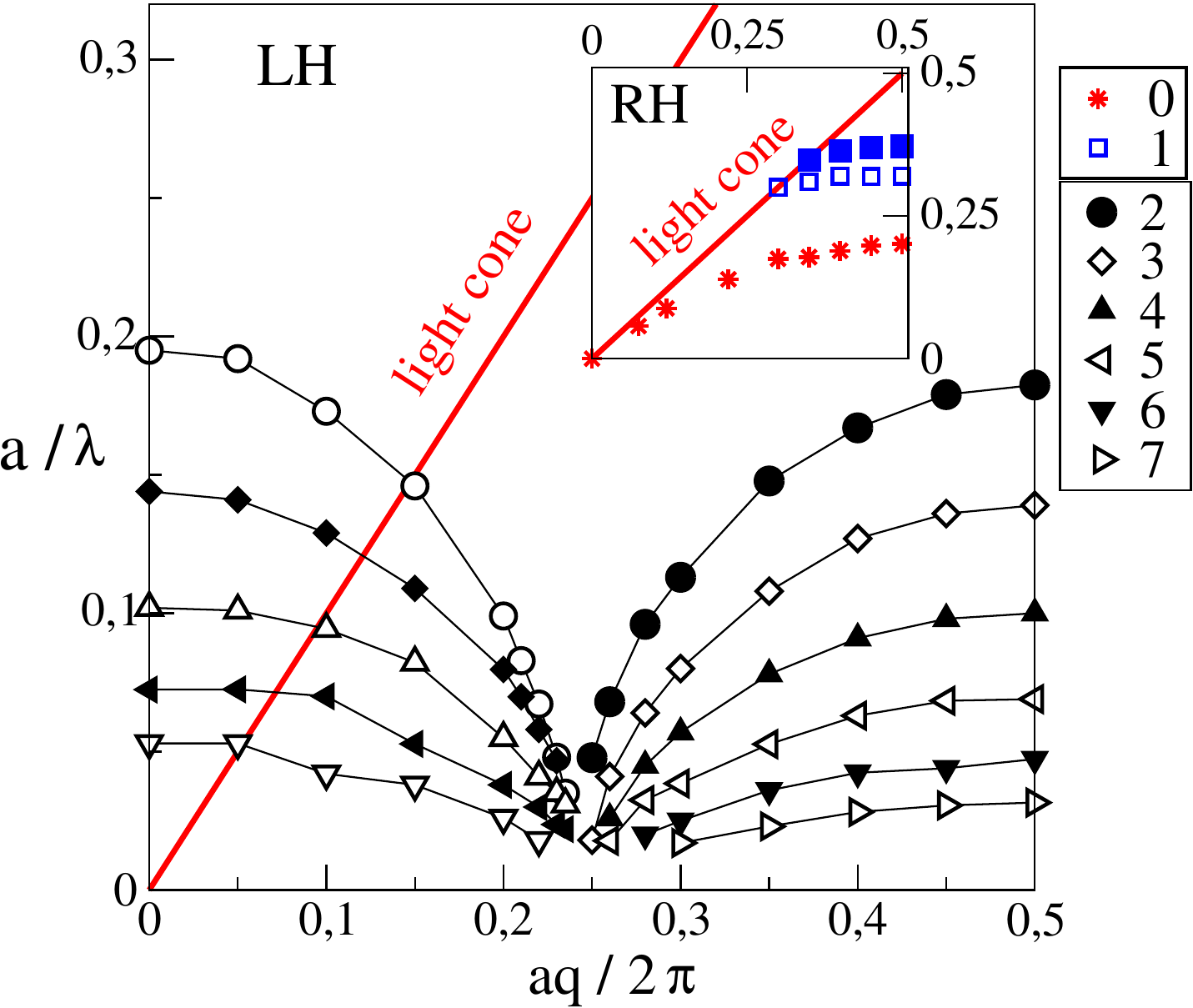}}
\caption{(Color online) Spectrum of guided  modes $f=a/\lambda = f(q)$ calculated for linear chain of LH cylinders
(Eq. \ref{eq:lin2}).  Open and full symbols correspond to spatially even and odd modes, respectively. Note that
modes change their spatial symmetry when $q$ crosses critical value $q_c\approx \pi/2a$.
For comparison with dielectric structures, the Inset shows the spectrum for linear chain of dielectric cylinders with permittivity $\varepsilon= +12$ and permeability $\mu=+1$.
}
\label{eps-12-gm}
\end{figure}

Resonances observed in the transmission spectra are given by excitations 
of eigenmodes of the photonic structures \cite{fan}.  
We calculate  the dispersion $f=f(q)$ of 
guided modes propagating along  the linear chain of LH cylinders   with wave vector $q$
and frequency $f=a/\lambda$. 
To find the dispersion curves $f(q)$, we  start with   two systems of linear equations 
for coefficients $\beta^+$ and $\beta^-$,  
(\ref{eq:lin2}) and (\ref{eq:lin3})
with zero right-hand side. Instead of solving these homogeneous systems of equations, we 
calculate the determinant as a function of frequency  \cite{economou}.
With the  use the Gauss-Jordan elimination method  \cite{nrcp} we    transform 
the matrices  in the l.h.s of Eqs.   (\ref{eq:lin2}) and  (\ref{eq:lin3})  into diagonal form
and  plot the frequency dependence of  inverse of obtained diagonal matrix elements. 
This enables us not only to find   the eigenfrequency and lifetime of a given of guided mode
but also to identify its  order $k$.   
Note that method enables us to calculate both guided modes with $f<q$ and leaky modes ($f>q$)  with finite lifetime \cite{economou}.

Figure \ref{eps-12-gm}  presents the dispersion curves $f=f(q)$  of guided modes.
For comparison, the spectrum of guided modes of an array of dielectric cylinders is shown in the Inset. As expected,
the spectrum  for left-handed cylinders is more complicated. 
In contrast to dielectric cylinders \cite{pm-pra}, the number of guided modes  is much higher.
The number of modes depends on the model parameters and increases when absolute value of the refractive index increases (data not shown). 
In agreement with Eq. (\ref{kk}),  their eigenfrequencies decreases when the mode index $k$ increases.

Another unexpected property of spectra of guided modes is the existence of  ''critical value'' of the wave vector $q_c$
($q_c\approx   \pi/2a$ in our model) for which no guided mode exists (Figure \ref{eps-12-gm}). 
Analysis of  another  left-handed structures with $\mu=-1$ (data not shown)  indicates that
$q_c$ depends neither on (negative)  permittivity nor on the radius of cylinder.

\section{Parameters of the model}

In this Section we investigate how the spectrum of resonant modes depends on the parameters of the model. 

\subsection{Radius of cylinder}

\begin{figure}[t]
\bc
(a)~~~~  R = 0.1 ~~~~~~~~~~~~~~~~~~~~~

{\includegraphics[width=0.78\linewidth]{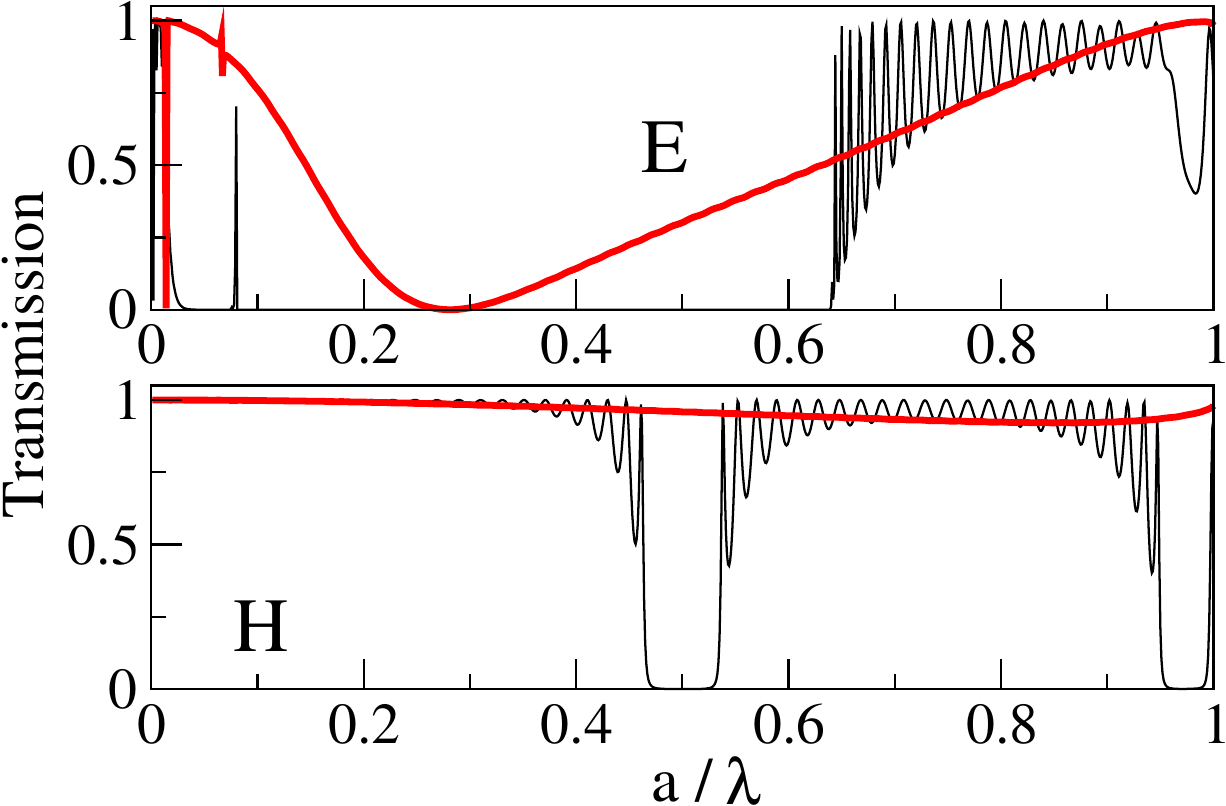}}\\
~~~\\

(b)~~~~  R = 0.2 ~~~~~~~~~~~~~~~~~~~~~

{\includegraphics[width=0.78\linewidth]{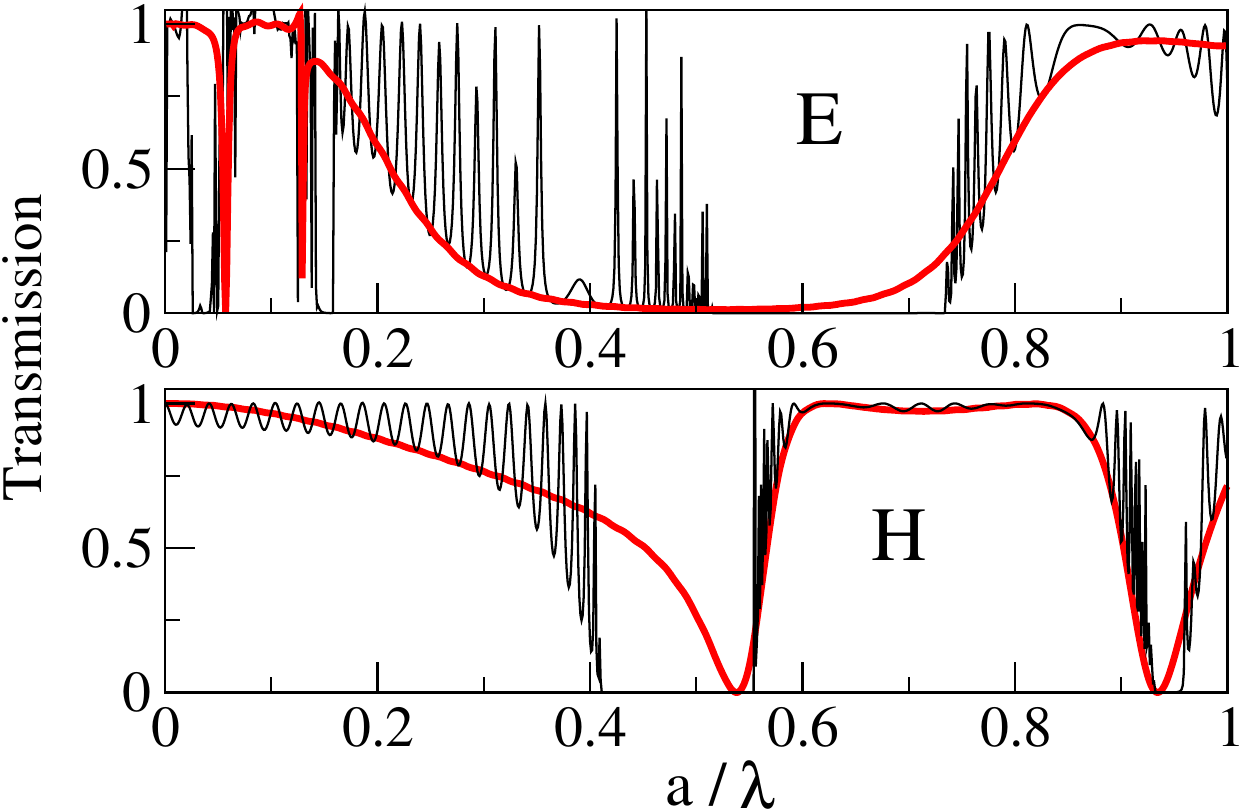}}

\footnotesize{
}
\caption{(Color online) Transmission coefficient of plane electromagnetic wave propagating across  the  linear array  of LH cylinders (bold red line)
and through periodic structure constructed from 24 rows of cylinders (thin black line). 
The radius of cylinders is $R=0.1 a$ (a) and $R=0.2 a$ (b).
Polarization of EM wave is given in panel legends.
}
\label{fig:r01T}
\end{center}
\end{figure}

Very  narow resonances in the transmission coefficient for the frequency  $a/\lambda< 0.08$ as well as broad minimum of the transmission at  $a/\lambda\sim 0.28$
shown in Fig. \ref{fig:r01T}(a) indicate that 
Fano resonances can be observed already in the linear chain of  very tiny 
($R=0.1a$) LH cylinders.  
The transmission coefficient of  photonic slab exhibits a broad gap for small frequencies with 
a few narrow resonances  
 at small frequencies $a/\lambda\le 0.09$. 
Broad transmission  band, typical for periodic structures, 
started at $a/\lambda \approx 0.65$. 

Figure \ref{fig:r01T}(b)  shows that the frequency dependence of the transmission coefficient is more dramatic when the radius of cylinders increases. For the $E_z$ polarized wave, the frequency dependence of $T$ is highly irregular,
indicating the high number of resonances in the structure. Only for higher frequency, regular transmission band 
is observed. 

Note that  in contrast to the $E_z$ waves,  the transmission coefficient for  the $H_z$ wave 
typical frequency dependence typical for  dielectric photonic crystals, 
independently on the   cylinder radius.
As we will see later, the absence of Fano resonances is due  to big difference 
in  absolute values of the permittivity $\varepsilon = -12$ inside cylinders and permittivity 
$\varepsilon_1=+1$ of the embedding medium.

\subsection{Permeability}

Photonic structures discussed up to now possess LH components with effective permeability $\mu=-1$  equal  exactly 
to minus value of the permeability $\mu_1$ of the embedding media. This coincidence might be responsible for strong resonances 
of electromagnetic field observed in previous Sections. We note that
spatial distribution of the  EM field with high values of the intensity 
close to cylinder boundary  (Fig. \ref{eps-12-pole}) is similar to 
 surface plasmons excited at the planar  interface of vacuum and LH media \cite{WP} and that
the dispersion relation of the
 $E_z$ polarized surface plasmon  wave (with electric field parallel to the interface) is 
singular when the permeability $\mu$ equals to minus value of permeability of embedding media
\cite{WP,vary}. 
Of course, the  analogy is not exact since, in contrast to our resonant states,  
surface plasmon  cannot be excited by incident EM wave. 

\begin{figure}[t]
\bc
\includegraphics[width=0.6\linewidth]{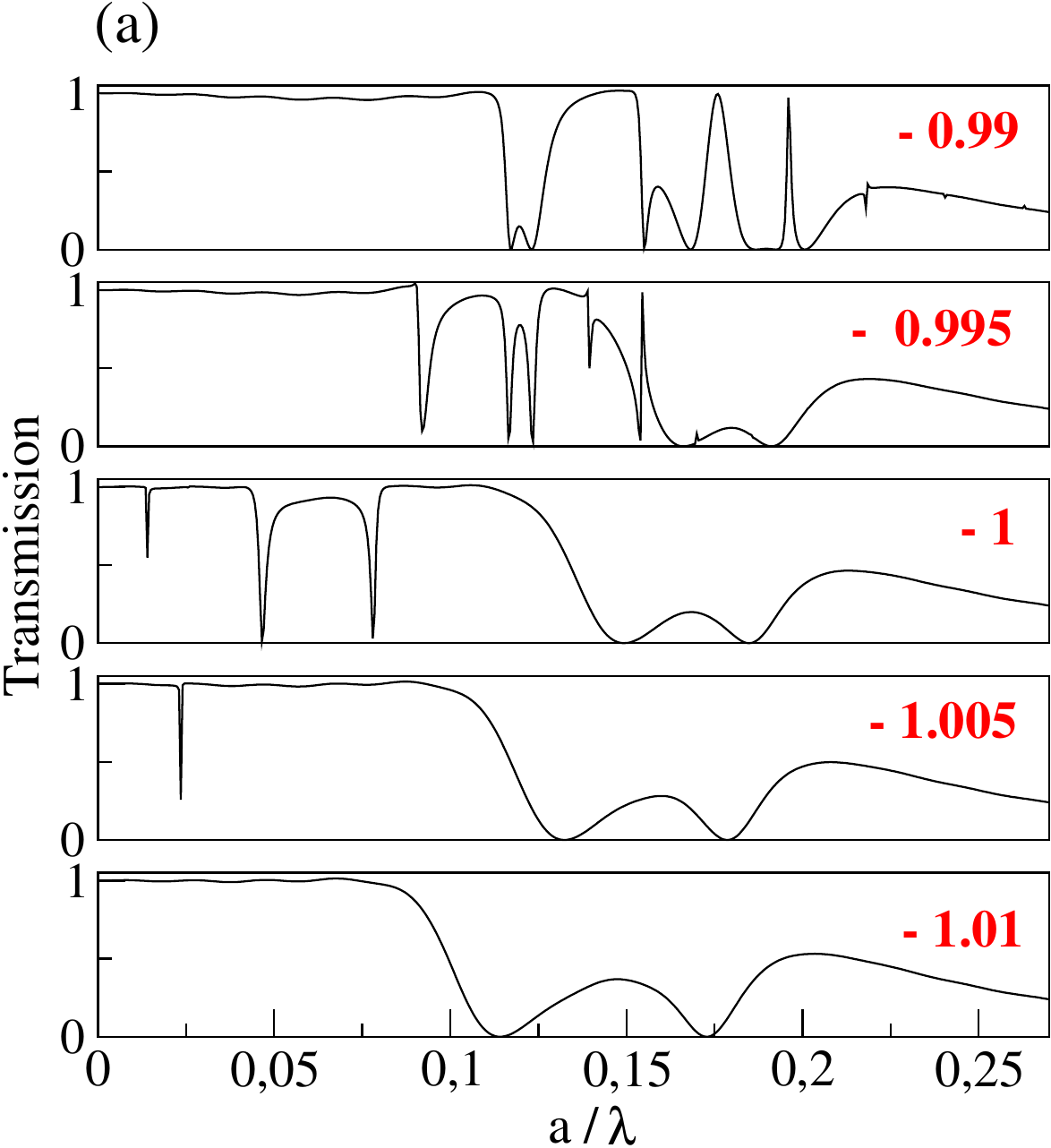}
~~~
\includegraphics[width=0.7\linewidth]{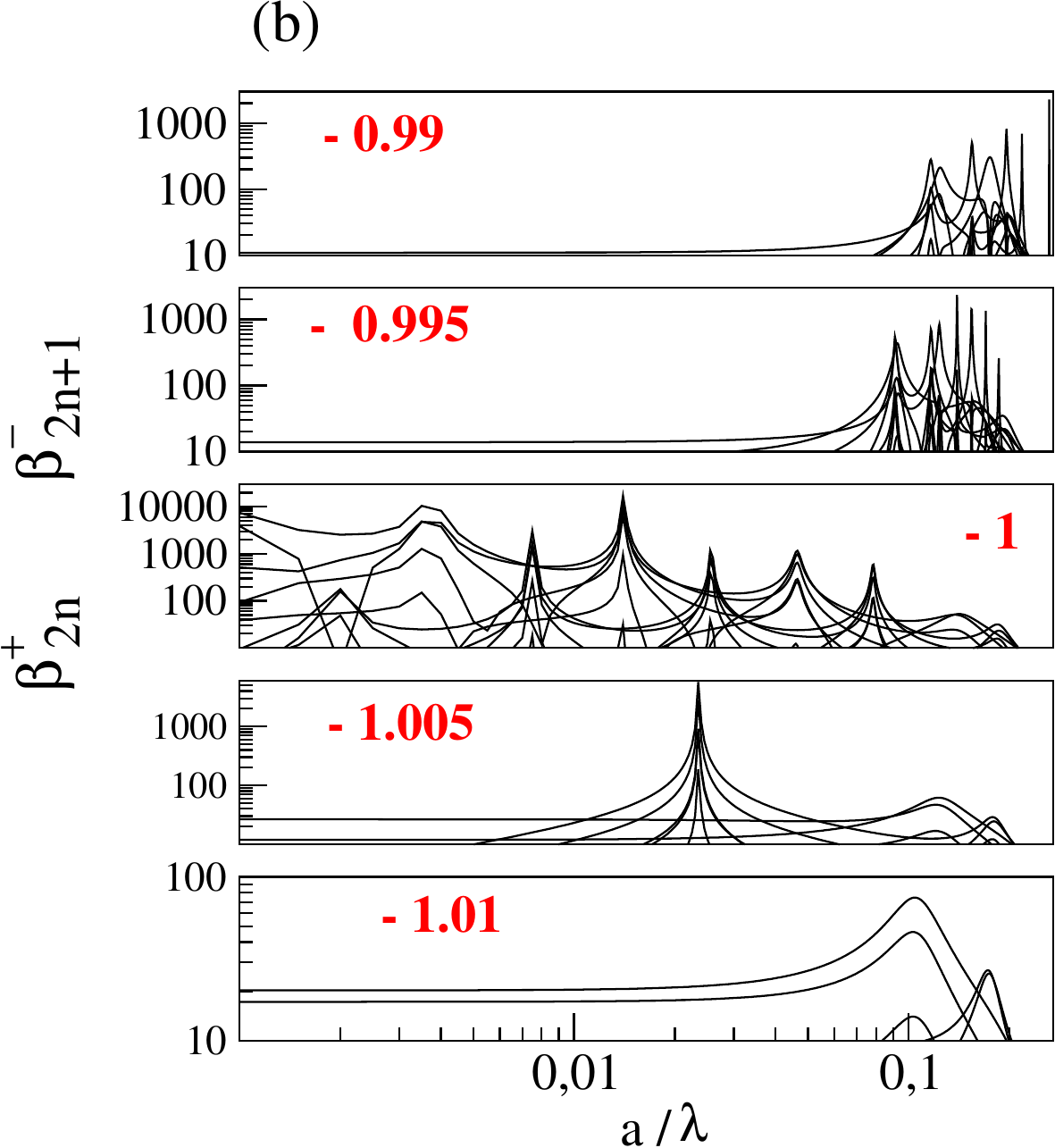}~~~~
\ec
\caption{(Color online)
(a)  Transmission coefficient $T$  for linear array  of LH cylinders
with radius $R=0.3a$, $\varepsilon=-12$ and various values of magnetic permeability $\mu$ (given in legends).
(b) Frequency dependence of coefficients $\beta$ for linear array  of cylinders.
Note the strong sensitivity  of resonant frequencies to small changes of permeability. 
Cylinders are embedded in vacuum with permeability $\mu_1=1$.
}
\label{mu-1}
\end{figure}

Figure \ref{mu-1}  shows the transmission coefficients and spectra of coefficients $\beta$ for linear chain of LH cylinders with
permeability slightly  below and above the critical value  $\mu = -1$. The positions of resonances in frequency spectra are  very sensitive to
the actual value of $\mu$: for smaller (in absolute value) $\mu$, resonances shift to frequencies $f>0.1$, while for larger values of permeability  they  disappear.    

\subsection{Dispersive LH materials}

\begin{figure}[t]
\begin{center}
\includegraphics[width=0.4\textwidth]{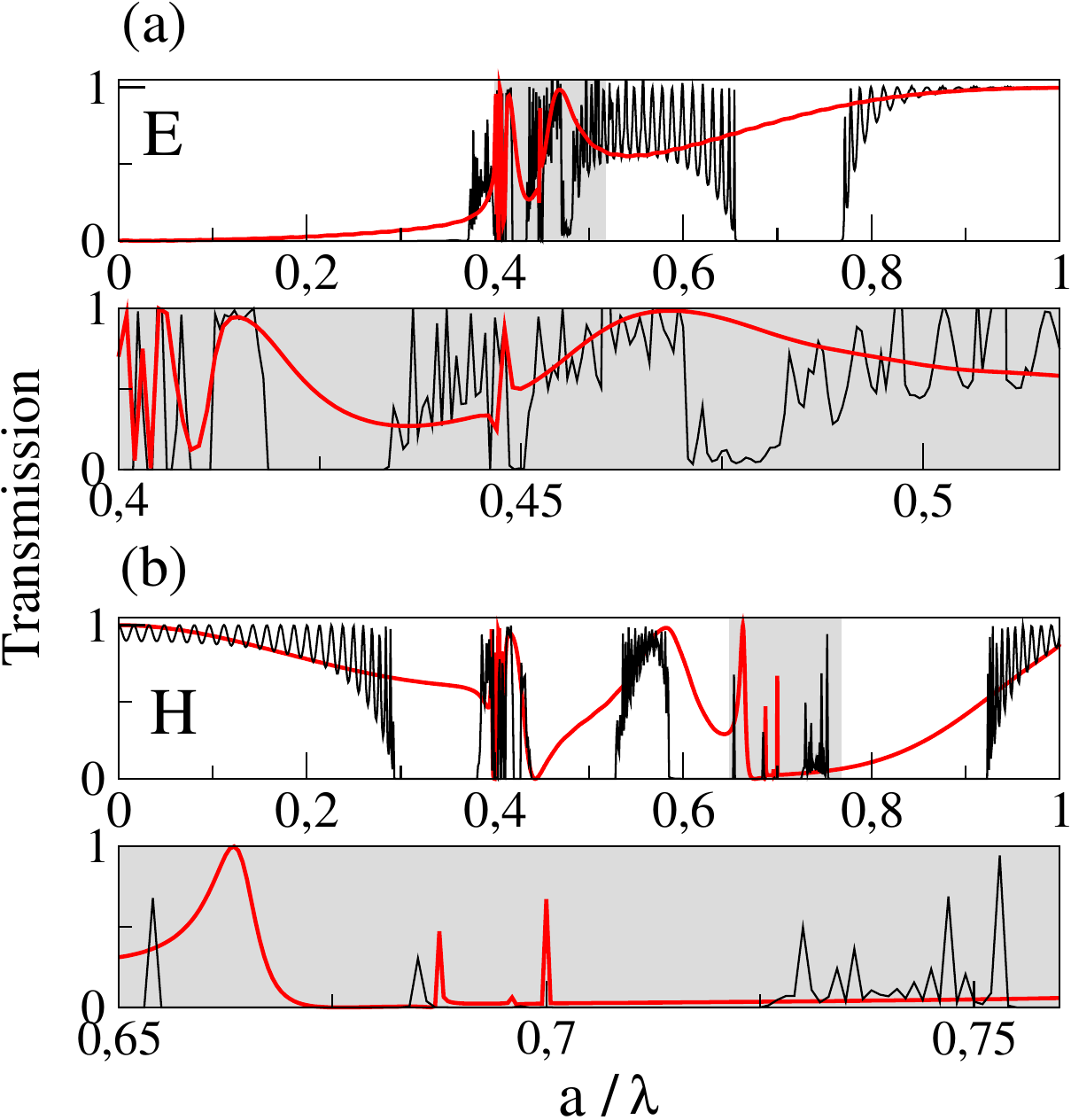}
\end{center}
\caption{(Color online) Transmission coefficient of plane electromagnetic wave propagating through square array of cylinders made from left-handed dispersive medium. 
The  cylinder permittivity and permeability is   given by Eqs. 
(\ref{eq:eps}) and 
(\ref{eq:mu}), respectively.
The permittivity  is negative in the whole frequency interval  $0<f<1$, and permeability 
is negative when  $f_0<f<f_0/\sqrt{0.6}= 0.516$.
(a)  $E_z$ polarization (intensity of electric field is parallel to the cylinder axis).
The transmission coefficient exhibits irregular frequency dependence of 
in the vicinity of frequency $f_e=a/\lambda_e=0.447$  where $\mu(f_e)\approx -1$.
For the $H_z$ polarized wave (panel (b)),
similar irregularities are observed around  the frequency $f_m=a/\lambda_m\approx 0.7$, where $\varepsilon(f_m)\approx -1$. 
}
\label{fig:lhm}
\end{figure}

Since  realistic LH materials  are dispersive \cite{veselago}, it is important  to generalize our 
analysis  to
structures made from  dispersive LH materials. 
We consider  LH cylinders
with frequency dependent permittivity
\be\label{eq:eps}
\varepsilon(f) = 1 - \frac{1}{f^2}
\ee
and  permeability 
\be
\label{eq:mu}
\mu(f) = 1 - 0.4 \frac{f^2}{f^2-f_0^2}~~~~~~~~(f_0=0.4).
\ee
Transmission coefficient for arrays of such cylinders  
is shown in   Fig. \ref{fig:lhm} both for the $E_z$ and $H_z$ polarized  EM waves.

For small frequency $f<f_0$ the permittivity is negative and permeability is positive.
The transmission coefficient  exhibits the frequency dependence typical for an array of  
metallic cylinders \cite{pendry-j} with low frequency transmission gap for the $E_z$ polarized wave
and regular transmission band for the $H_z$ polarized wave. Irregular frequency dependence 
for $f$ slightly above $f_0=0.4$ is due to  the large value of the effective index  LH material.
More interesting is the transmission coefficient for 
the $E_z$ polarized wave in the frequency region around
$f_e = 0.447$ defined by relation $\mu(f_e)=-1$   which 
exhibits irregular frequency dependence similar to that discussed in Sect. 2. Similar resonant behavior is observed
for the  $H_z$-polarized wave in the vicinity of frequency $f_m=0.7$, where the permittivity of LH material approaches minus unity.

\subsection{Absorption}

\begin{figure}[t]
\centering
{\includegraphics[width=0.7\linewidth]{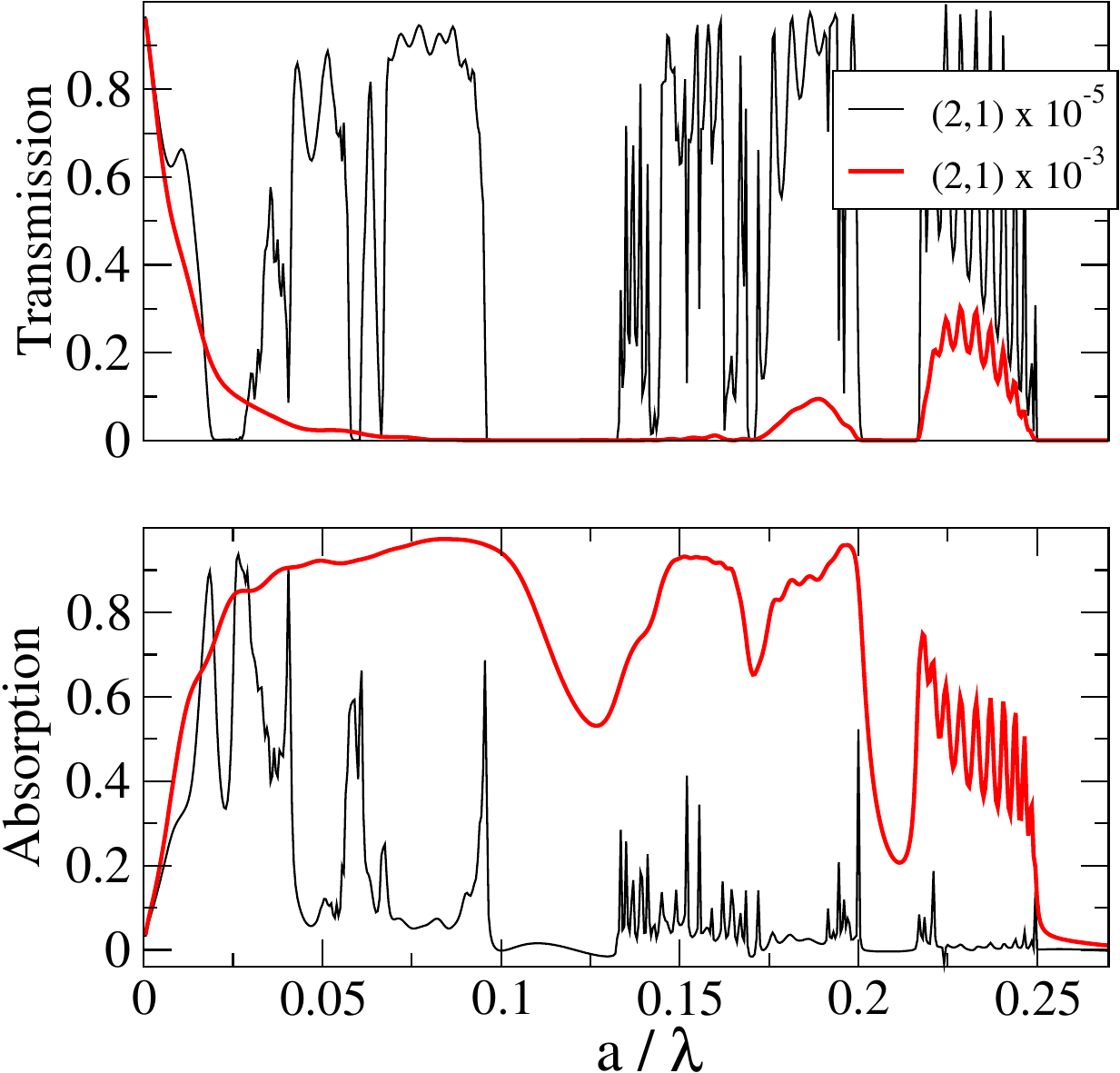}}
\caption{(Color online) Transmission coefficient and absorption for the photonic crystal with cylinders made from the left-handed material 
($\varepsilon=-12$, $\mu=-1$) and with small imaginary part of permittivity and permeability (given in legend) as a function of the dimensionless frequency $a/\lambda< 0.27$. 
}
\label{eps-12-straty-xasa}
\end{figure}

As discussed in \cite{pm-pra}, high intensity of EM field inside cylinders may cause strong  enhancement of the absorption 
of transmitted EM wave for frequencies close to resonant Fano frequency.
Figure \ref{eps-12-straty-xasa} demonstrates an increase of the absorption in photonic structures with LH cylinders. 
Absorption is close to unity already for very small values of imaginary part of permittivity and permeability,
Imag $\varepsilon,~\mu\sim 10^{-3}$.
Since Fano resonances lye close to each other, we obtained  broad absorption band 
with  absorption coefficient close to unity.

\section{Conclusion}

We have shown that transmission spectra of photonic structures with LH components might be strongly influenced by Fano resonances
excited by incident electromagnetic wave. Fano resonances play an  important  role 
especially when the permittivity and/or permeability of the LH material is close to $-1$. 
As such values exists  in any 
realistic dispersive LH medium,   we expect  that irregular resonant behavior, 
discussed in this paper, might  be observed experimentally.
However,  non-zero imaginary part of permittivity and permeability strongly enhanced absorption losses
due to large amplitudes of electromagnetic waves inside cylinders. 
Therefore, instead of irregular  resonant transmission,  
found  numerically in   loss-less materials, we expect that a broad   absorption band in the resonant frequency region will be measured.

We also found that the spatial distribution of the electromagnetic  field inside the cylinders is highly inhomogeneous with strong field enhancement  in the narrow region in the vicinity of  the boundary of cylinder. 


\bigskip

This work was supported by the Slovak Research and Development Agency under the contract No. APVV-0108-11
and by the Agency  VEGA under the contract No. 1/0372/13.

\appendix

\section{Numerical methods}

In this Section we describe the numerical method used for the calculation of the transmission of electromagnetic wave 
with wavelength $\lambda$ propagating across  linear chain of cylinders shown in Fig. 1(a).
Owing to the cylinder  symmetry, we express electric and magnetic fields in cylindrical coordinates.
Consider  the $E_z$-polarized electromagnetic wave, $\vec{E}=(0,0,e_z)$ (the $H_z$ polarized wave an be treated  in the same way). 
Two non-zero  components of
the magnetic field, the angular, $h_\phi$, and the radial, $h_r$, ones  can be found  with the use 
of    Maxwell equations
\cite{jackson} 
\be\label{app:1}
 i\omega\mu h_\phi(r,\phi) = - \frac{\partial e_z}{\partial r},~~~~~
i\omega\mu h_r(r,\phi) =  \frac{1}{r}\frac{\partial e_z}{\partial \phi}.
\ee
For 
cylinder centered in $(x,y) = (0,0)$ we  express 
the electric intensity $e_z$  as a  sum  of cylinder functions
\cite{stratton}. Inside the cylinder ($r\le R$) we have 
\begin{equation}
\label{eq:inc}
\begin{array}{rclcl}
e_z^{\rm in}(r,\phi) = J_0(v)\alpha_0^+ 
&\!\!\!\!+& 2\displaystyle{\sum_{k=1}^{} }\alpha_k^+J_k(v) \cos(k\phi)\\ 
&\!\!\!\!+& 2i\displaystyle{\sum_{=1}^{} }\alpha_k^-J_k(v) \sin(k\phi), 
\end{array}
\end{equation}
where $J_k$ is  Bessel function of integer order,
 $v=2\pi rn/\lambda$,
and $n=\sqrt{\varepsilon\mu}$ is the 
index of refraction.
Coefficients   $\alpha^+$ and $\alpha^-$ determine amplitudes of even ($\propto \cos k\phi$) and odd ($\propto\sin k\phi$)
cylindrical waves, respectively.
In Eq. (\ref{eq:inc}), as well in all expressions below, the sumation over $k$ is restricted to $N_B$ lowest order cylinder functions ($k\le N_B$).

The electric field outside the cylinder  consists from  three contributions: the first one is the field  scattered 
on  the cylinder itself \cite{stratton}
\begin{equation}
\label{eq:onc}
\begin{array}{lcl}
e_z^0(r,\phi) = H_0(u)\beta_0^+ 
&\!\!\!\!+& 2\displaystyle{\sum_{k=1}^{}} \beta_k^+H_k(u) \cos(k\phi)\\
&\!\!\!\!+&
 2i\displaystyle{\sum_{k=1}^{}} \beta_k^-H_k(u) \sin(k\phi)\\
(u=2\pi r/\lambda). &
\end{array}
\end{equation}
($r\ge R$)
Here,
$H_k(z) = J_k(z) + i Y_k(z) $ is the first Hankel function
\cite{stratton,AS}.

\begin{figure}[t]
\begin{center}
\includegraphics[width=0.3\textwidth]{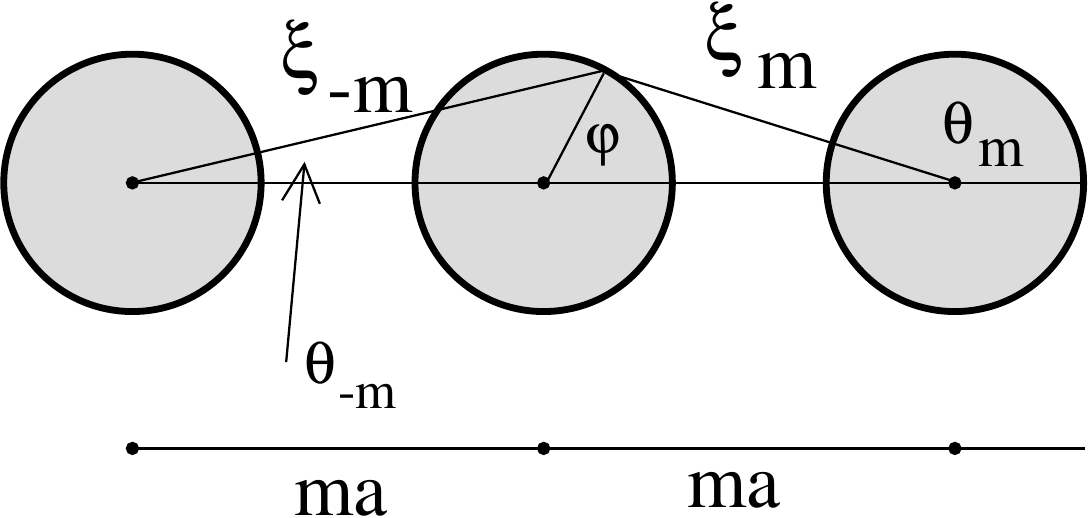}
\caption{Parameters used in derivation of Eq. (\ref{eq:m}).
}
\label{fig:gegen_n}
\end{center}
\end{figure}

The second contribution to the external fields consists from  fields scattered on other cylinders 
present in the structure.
In particular, the contribution of the $m$th cylinder
 ($m=\pm 1, \pm 2,\dots ,N_s$) expressed as a function of
coordinates associated with 
its  center (Fig. \ref{fig:gegen_n})
reads
\begin{equation}\label{eq:m}
\begin{array}{ll}
e_z^m(\xi_m,\theta_m) =& H_0(w_m)\beta_{m0}^{+} \\
&+ 2\displaystyle{\sum_{k>0}} \beta_{mk}^+H_k(w_m) \cos(k\theta_m)\\
 &+ 2i\displaystyle{\sum_{k>0}} \beta_{mk}^-H_k(w_m) \sin(k\theta_m)\\
 (w_m = 2\pi \xi_m/\lambda).   &  
\end{array}
\end{equation}
Note that coefficients $\beta_{mk}$ differs from $\beta_k$ defined in Eq. (A.3). However, 
thanks to the periodicity of the structure  along the  $x$  axis, 
we can use the Bloch's theorem and express  $\beta_{mk}^\pm$ of the $m$th cylinder 
in terms of coefficients $\beta_k^\pm$ as 
\begin{equation}
\beta_{mk}^\pm = e^{iqma}\beta_k^\pm
\end{equation}
where $q$ is 
the $x$-component of the  wave vector of EM field.

Finally, the third contribution is given  by incident electromagnetic wave. For instance,
the plane wave incident perpendicularly to the cylinder chain (Fig. 1(b)) can be expressed as a sum
of Bessel function \cite{AS}
\begin{equation}
\label{eq:ini}
\begin{array}{rl}
e^i_z(r,\phi) = J_0(u) 
&\!\!\!\!\!+ 2 \displaystyle{\sum_{k=1}}J_{2k}(u)  \cos(2k\phi)  \\
 &+ 2i \displaystyle{\sum_{k=1}} J_{2k-1}(u) \sin[(2k-1)\phi].
\end{array}
\end{equation}

\medskip

Unknown coefficients $\alpha$ and $\beta$
defined in Eqs. (A.2-A.4)  are  calculated from the  requirement of the  
continuity of tangential components of electric and magnetic fields  at the boundary of each cylinder.
Thanks to Eq. (A.5), it is sufficient to formulate the continuity conditions only for the cylinder located  at 
the center of coordinates. 
Since 
  $e_z$ 
is parallel to cylinder surface, the requirement of the  continuity of electric field is easy to formulate,
\begin{equation}
\label{eq:lin-e}
\begin{array}{ll}
e_z^{\rm in}(R^-,\phi) &\!\!\!= e_z^{\rm out}(R^+,\phi)\\
  ~ & ~ \\
 &\!\!\!= e_z^i(R^+,\phi) +e_z^0(R^+)
 + \sum_{n\ne 0} e_z^n(\xi_n,\theta_n).
\end{array}
\end{equation}
The continuity condition for the  magnetic field is expressed in more complicated form
\begin{equation}
\label{eq:lin-h}
\begin{array}{ll}
&h_\phi^{\rm in}(R^-) = h_{\rm tang}^{\rm out} = h^i_\phi(R^+,\phi) + h_\phi^0(R^+,\phi)\\
& + \sum_{n\ne 0} \left[ h_{\theta_n}^n(\xi_n,\theta_n)\cos \alpha_n  - h^n_{r_n}(w_n,\theta_n)\sin \alpha_n \right]
\end{array}
\end{equation}
where the angle $\alpha_n = \theta_n-\phi$.

\begin{figure}[t]
\begin{center}
\includegraphics[width=0.3\textwidth]{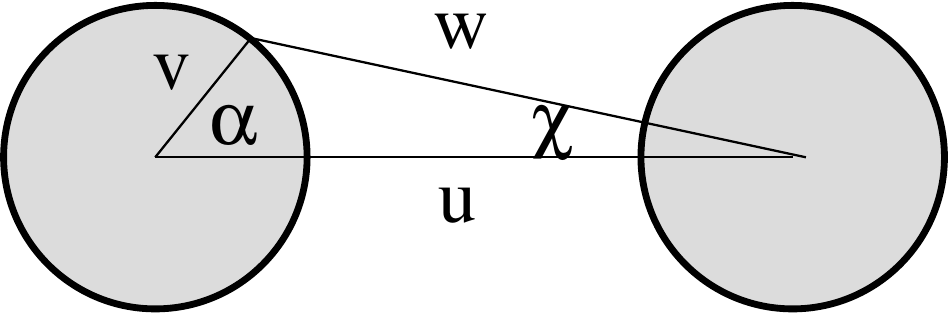}
\caption{Parameters  used in  Gegenbauer's relation, equations (\ref{eq:geg}) and  (\ref{eq:gegen}).
}
\label{fig:gegen}
\end{center}
\end{figure}

Before  solving  this  system of equations, we  express all fields  in Eq.  (\ref{eq:m}) 
in terms of variables   $r$ and $\phi$ associated with the cylinder located at 
the point $x=0$, $y=0$. This can be done 
with the use of the Gegenbauer formula for cylindrical functions (Fig.  \ref{fig:gegen})
\begin{equation}\label{eq:geg}
H_m(w) e^{\pm im\chi} = \sum_{k=-\infty}^{+\infty}  H_{m+k}(u)J_k(v)e^{\pm ik\alpha}
\end{equation}
and their  derivative $H'_m(w)$: 
\begin{equation}
\label{eq:gegen}
H'_m(w) e^{\pm im\chi} = \sum_{k=-\infty}^{+\infty}  H'_{m+k}(u)J_k(v)e^{\pm ik\alpha}
\end{equation}
\cite{AS}. 
Inserting (\ref{eq:gegen}) into  Eq. (\ref{eq:m}) 
we express, after some algebra,  the fields at the outer boundary of the cylinder in the form
\begin{equation}
\label{eq:continuity}
e_z^{\rm out}(R^+,\phi) = 
\displaystyle{\sum_{k,m=0}^{}} \textbf{B}_{km}\beta^+_m\cos k\phi + 
\displaystyle{\sum_{k,m=1}^{}}\textbf{C}_{km}\beta^-_m\sin k\phi
\end{equation}
and
\be
h_t^{\rm out}(R^+,\phi) =
\displaystyle{\sum_{k,m=0}^{}} \textbf{B'}_{km}\beta^+_m\cos k\phi +
\displaystyle{\sum_{k,m=1}^{}} \textbf{C'}_{km}\beta^-_m\sin k\phi.
\ee
Explicit expressions for   matrices
$\textbf{B}$,
$\textbf{B'}$,
$\textbf{C}$ and
$\textbf{C'}$
are given in  Appendix B.

In the next step, we eliminate coefficients $\alpha$ from Eqs. (A.11) and (A.12) and obtain two separate systems of linear equations for coefficients $\beta^+$ and $\beta^-$:
\begin{equation}
\label{eq:lin2}
\sum_m \left[\textbf{B}_{km} - \zeta\frac{{\cali}_k}{{\cali}'_k}\textbf{B'}_{km}\right]\beta^+_m = 
e_k^+ - \zeta\frac{{\cali}_k}{{\cali}'_k} h_k^+, 
\end{equation}
$k,m  = 0,1,\dots N_B$,  and
\begin{equation}
\label{eq:lin3}
\sum_m \left[\textbf{C}_{km} - \zeta\frac{{\cali}_k}{{\cali}'_k}\textbf{C'}_{km}\right]\beta^-_m = 
e_k^- - \zeta\frac{{\cali}_k}{{\cali}'_k} h_k^-, 
\end{equation}
$k,m = 1,\dots N_B$.
Here,
${\cali}_k = J_k(2\pi nR/\lambda)$ and
\be
\zeta = \displaystyle{\sqrt{\frac{\mu}{\varepsilon}}} ~~~~(\textrm{Real}~\zeta >0)
\ee
is the impedance of cylinders.
In numerical analysis,
we consider the number of cylinders $N_s\le 20 000$ and number of modes $N_B\le 12$. 

\medskip

Transmission coefficient can be calculated as the ratio of the $y$-component of the Poynting vector $S_x(y_p)$,
calculated for any $y_p>r$
to the incident Poynting vector, $S_y^i$
\begin{equation}
\label{eq:T}
T = \frac{S_x(y_p)}{S_x^i}
\ee
where
\be 
S_x(y_p) = 
\int_{-a/2}^{+a/2} {\rm d} x~ e_z(x,y_p) h^*_x(x,y_p)
\end{equation}
and
\be
{S_x^i} = 
\int_{-a/2}^{+a/2} {\rm d} x~ e^i_z(x,y_p) (h^i_x(x,y_p) )^*.
\ee
Similarly, reflection coefficient $R$ is given by
\begin{equation}
\label{eq:R}
R = \frac{S_x(y_p)}{S_x^i}
\ee
Where
\be 
S_x(y_p) = 
\int_{-a/2}^{+a/2} {\rm d} x~ (e_z-e_z^i) (h_x-h^i_x)^*
\end{equation}
calculated for  $y_p<-r$.

\section{Matrices \textbf{B} and \textbf{C}}

With notation
$J_k \equiv J_k(u)$
and ${\cal H}_k = H_k(u)$, $u=2\pi R/\lambda$ we
express the explicit form of
the $(N_B+1)\times (N_B+1)$ matrix \textbf{B} 
\begin{equation}\label{eq:B}
\begin{array}{lcl}
B_{00} &=& {\calh}_0 + 
\displaystyle{\sum_{n=1}^{N_s}}~~
2\cos q H_0(un)J_0\\
B_{0m} &=& 
\displaystyle{\sum_{n=1}^{N_s}}~~
H_m(un)J_0 \times [(-1)^m e^{iqan}+ e^{-iqan} ]\\
B_{k0} &=& 
\displaystyle{\sum_{n=1}^{N_s}}~~
H_k(un)J_k \times 2[ e^{iqan} + (-1)^ke^{-iqan} ] \\
  && \\
B_{km} &=& 
2{\calh}_k\delta_{km} + 
\displaystyle{\sum_{n=1}^{N_s}}~~
[e^{iqan}(-1)^{m-k} + e^{-iqan}]\\
&&[H_{m-k}(un)+(-1)^{k}H_{m+k}(un)] J_k
\end{array}
\end{equation}
($k,m= 1,2,\dots N_B$).  
The $N_B\times N_B$ matrix \textbf{C} has a form
\begin{equation}
\begin{array}{ll}
\label{eq:C}
C_{km} = 2{\calh}_k\delta_{km} + 
\displaystyle{\sum_{n=1}^{N_s}}&
[e^{iqan}(-1)^{m-k} + e^{-iqan}]\times \\
& [{H}_{m-k}(un)-(-1)^{k}{H}_{m+k}(un)] J_k
\end{array}
\end{equation}
Matrices \textbf{B'} and \textbf{C'} could be obtained from \textbf{B} and \textbf{C}, respectively, by substitutions 
\begin{equation}
J_k \to J'_k, ~~~~~{\calh}_k\to{{\calh}'}_k.
\end{equation}

\section{System with $N$ rows of cylinders}

If the system consists of $N$ rows of cylinders 
(the $n$th row lies in plane $y=(n-1)a,z$), then 
we have to define $N$ sets of parameters $\beta^\pm_{yk}$, each for the $n$th cylinder along the $y$ direction.
The method of calculation remains the same as for the linear chain with $N=1$, but  resulting linear
relations between  coefficients
$\beta^\pm_{nk}$ are more complicated. In particular, it is not possible to separate coefficients $\beta^+$ and
$\beta^-$.
Resulting system of linear equations for coefficients $\beta^\pm_{nk}$ is of the size $N\times (2N_B+1$).
For $N_B=12$ (typically used in numerical simulations) and $N=24$ the problem reduces to the solution of 600 
linear equations.

\end{document}